\documentclass[aps,prl,twocolumn,a4paper,superscriptaddress]{revtex4}%
\usepackage{amsfonts}
\usepackage{amsmath}
\usepackage{amssymb}
\usepackage{graphicx}%
\DeclareGraphicsExtensions{.eps,.pdf}
\providecommand{\U}[1]{\protect\rule{.1in}{.1in}}
\ifx\pdfoutput\relax\let\pdfoutput=\undefined\fi
\newcount\msipdfoutput
\ifx\pdfoutput\undefined\else
\ifcase\pdfoutput\else
\ifx\paperwidth\undefined\else
\ifdim\paperheight=0pt\relax\else\pdfpageheight\paperheight\fi
\ifdim\paperwidth=0pt\relax\else\pdfpagewidth\paperwidth\fi
\fi\fi\fi
\begin{document}
\preprint{HEP/123-qed}
\title[Short title for running header]{Dispersion and damping of multi-quantum well polaritons from resonant
Brillouin scattering by folded acoustic modes}
\author{B.\ Jusserand}
\affiliation{Institut des Nanosciences de Paris, CNRS UMR7588, UPMC, 75005 Paris, France}
\author{A.\ Fainstein}
\affiliation{Instituto Balseiro and Centro atomico Bariloche, CNEA, R8402AGP Bariloche, Argentina}
\author{R.\ Ferreira}
\affiliation{Laboratoire Pierre Aigrain, ENS, CNRS UMR 8551, UPMC, UPD, 75005 Paris, France}
\author{S.\ Majrab}
\affiliation{Institut des Nanosciences de Paris, CNRS UMR7588, UPMC, 75005 Paris, France}
\author{A.\ Lemaitre}
\affiliation{Laboratoire de Photonique et de Nanostructures, CNRS UPR20, 91460 Marcoussis, France}

\begin{abstract}
We report on confined exciton resonances of acoustic and folded acoustic
phonon light scattering in a GaAs/AlAs multi-quantum-well.\ Significant
variations of the line shifts and widths are observed across the resonance and
quantitatively reproduced in terms of the polariton dispersion.\ This high
resolution Brillouin study brings new unexpectedly detailed informations on
the polariton dynamics in confined systems.

\end{abstract}
\startpage{1}
\endpage{2}
\maketitle

Resonant Brillouin scattering has played a very important role in the
experimental validation of the concept of exciton polariton in bulk
semiconductors and has provided the first determination of the specific energy
dispersion of these excitations close to the exciton transition
\cite{weisbuch1982}. In such a case, strong dispersion effects appear in the
dielectric constant, significantly affecting the relation between the energy
and the wavevector of the electromagnetic excitations. Based on the linear
dispersion of acoustic waves in solids, a direct relation exists between the
photon energy shift measured by Brillouin scattering and the dispersion in the
compound of the incoming and outgoing polaritons involved in the scattering
event. Strong variations of the Brillouin shift are then expected from the
wavevector and energy conservation in the presence of dispersion. These
effects have been reported in several bulk semiconductors and the energy
variation was used to determine the polariton dispersion close to the
fundamental exciton transition. When the finite lifetime of the exciton is
taken into account, the dielectric constant becomes a complex quantity in the
vicinity of the exciton energy and polariton damping is expected to emerge.
The related variation of the Brillouin linewidth has escaped up to now the
observation because of the limited resolution of standard Raman setups.
Additional complications appear in the polariton picture in bulk materials
because of the exciton energy dispersion: two polariton branches coexist at
the same energy in some cases and the conversion of a single photon outside
the sample into these multiple branches has never been completely understood
\cite{bendow1978}. More recently, the concept of exciton polariton has been
extensively reconsidered in the context of microcavities \cite{weisbuch1992}.
In these photonic monolithic resonators, two-dimensional cavity photons with
dispersion in the plane of the cavity strongly interact with two-dimensional
excitons confined in quantum wells inserted in the photonic cavity, leading to
a large number of novel optoelectronic properties. Inelastic light scattering
has been applied to demonstrate the relevance of the polaritonic picture to
describe light scattering by cavity polaritons \cite{fainstein1997}. It has
not been possible however to follow the light scattering spectra across the
full excitonic resonance contrary to what was previously done for bulk
polaritons. Moreover Brillouin studies with a determination of incoming energy
resolved energy shifts are not available up to now. We will show in this
letter that these issues can be conclusively addressed using high resolution
Brillouin studies of high quality multi-quantum wells.

Surprisingly, semiconductor superlattices have attracted very little attention
from the point of view of Brillouin scattering close to strong excitonic
resonances, while being extensively studied for their unique optical
properties. Moreover, superlattices exhibit acoustic phonon folding
\cite{jusserand1989} due to new periodicity associated with the periodic
alternance of semiconductors with contrasting acoustic and acousto-optic
properties. This new feature opens the route for the investigation of exciton
Brillouin resonances using several phonon probes with different energies,
instead of the unique low energy Brillouin active acoustic phonon in bulk
materials. A series of reports has been focused on the emergence at strong
excitonic resonances of dominant contributions to the Brillouin signal from
isolated quantum wells due to fluctuations of well widths from well to well
\cite{ruf1993}. The activation of the full density of acoustic states results
from the localization of the intermediate states involved in the scattering
process, while no evidence is obtained of polariton related resonant features
on wavevector conserving acoustic phonons.

In multi-quantum wells, a type of superlattices in which the barriers are
sufficiently thick to make negligible the coupling between electron states
confined in neighboring quantum wells, excitons no longer display any
significant dispersion along the growth axis \cite{andreani1994}. This
provides a unique realization of the simple theoretical picture
\cite{hopfield1958} of a discrete level in strong interaction with a photonic
continuum in experimental situations where the in plane wavevector is fixed to
zero by the geometry. In this case and taking into account excitonic damping,
the polaritonic and the excitonic pictures merge together into a common
description with a single excitation branch with significant dispersion and
damping close to the excitonic energy. This excitation is well described as a
photon dressed by a dispersive complex dielectric constant.

We report in this letter Brillouin scattering by acoustic (LA) and folded
acoustic (FLA) longitudinal modes in a GaAs/AlAs multi-quantum well. We have
been able to follow the variations across the excitonic resonances of the
energy, the width and the intensity of several well defined acoustic branches
showing that a well defined phonon wavevector is preserved throughout the
whole energy range. Thanks to the highest energy of the folded phonons,
dispersion of polariton energy and damping in the incoming and the outgoing
channels are well separated while there are mixed when the low energy
Brillouin line is considered. Consistent informations on the background
dielectric constant, on the excitonic oscillator strength and damping are
extracted from the different experimental observations on LA and FLA modes.

The eigenvalue problem for interacting photons and excitons has been first
solved for bulk materials by Hopfield \cite{hopfield1958} introducing new
coupled excitations called exciton polaritons. The polariton eigenvalue
equation writes:%
\[
\frac{c^{2}k^{2}}{\omega^{2}}=\varepsilon_{0}+\frac{4\pi\beta\omega_{X}^{2}%
}{\omega_{X}^{2}-\omega^{2}}%
\]
in which $\hbar ck$ is the photon energy and~$\hbar\omega_{X}$ the exciton
one. $\varepsilon_{0}$ is the background dielectric constant and $\beta$ a
number describing the strength of the exciton-photon coupling. This equation
leads to the existence of two branches of polaritons, the lower branch with
energies extending from zero to $\hbar\omega_{X}$ and the upper branch
extending from $\hbar\omega_{X}\sqrt{1+\frac{4\pi\beta}{\varepsilon_{0}}}$ to
infinity. Introducing $Q=\frac{4\pi\beta}{\varepsilon_{0}}$, the new
dispersion can be described by two parameters: a) the longitudinal transverse
splitting ($\omega_{LT}\simeq Q\omega_{X}/2$ for small values of $\beta$), the
width of the gap between the lower and the upper polariton branches in which
no states exist due to the polariton coupling and b) the Rabi splitting
($\omega_{R}\simeq\omega_{X}\sqrt{Q}$), the minimum separation between the two
branches at a given wavevector. The weight of the photon and exciton
components in the polariton wavefunctions strongly varies around the exciton
energy in a range given by $\omega_{R}$. Using standard values of the
polariton dispersion in GaAs \cite{weisbuch1982}, the dispersion parameters
read: $\omega_{LT}=0.086meV$ and $\omega_{R}=16meV$. Brillouin scattering is
strongly affected by the polaritonic coupling.
%TCIMACRO{\FRAME{ftbpFU}{3.1125in}{2.3929in}{0pt}{\Qcb{Schematic description of
%the resonant Brillouin process between an ingoing polariton (red circles) and
%outgoing polaritons (red triangles) for Stokes (ST) and anti-Stokes (AS)
%processes with the LA phonon (left panel) and the lowest folded phonon FLA-1.
%The solid lines (resp. dashed lines) show the real part and the imaginary part
%of the polariton wavevector deduced from Eq.(\ref{eq}). The parameters of the
%calculations are typical for the studied sample.}}{}{Figure}%
%maintain-aspect-ratio TRUE;  display "USEDEF";  valid_file "T";
%width 3.1125in;  height 2.3929in;  depth 0pt;  original-width 4.2393in;
%original-height 3.2517in;  cropleft "0";  croptop "1";  cropright "1";
%cropbottom "0";  tempfilename 'LKEK6R00.wmf';tempfile-properties "XPR";}}}%
%BeginExpansion
\begin{figure}
[ptb]
\begin{center}

\includegraphics[
natheight=2.392900in,
natwidth=3.112500in,
height=2.3929in,
width=3.1125in
]%
{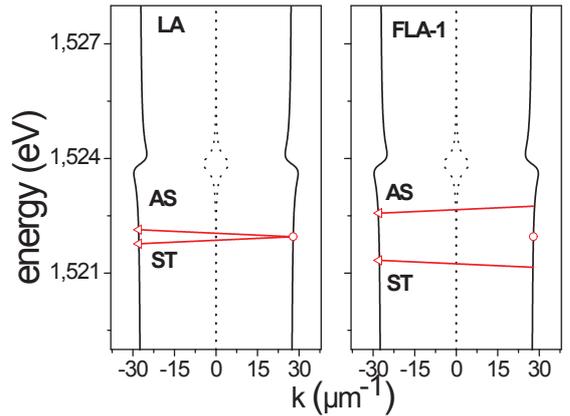}%
\caption{Schematic description of the resonant Brillouin process between an
ingoing polariton (red circles) and outgoing polaritons (red triangles) for
Stokes (ST) and anti-Stokes (AS) processes with the LA phonon (left panel) and
the lowest folded phonon FLA-1. The solid lines (resp. dashed lines) show the
real part and the imaginary part of the polariton wavevector deduced from
Eq.(\ref{eq}). The parameters of the calculations are typical for the studied
sample.}%
\end{center}
\end{figure}
%EndExpansion
In the standard picture, the incoming photon is scattered into an outgoing one
with the emission of an acoustic phonon whose energy and wavevector is
obtained from the energy and wavevector conservation during the scattering
process. As sound velocities are much smaller than light velocities in solids,
the phonon wavevector equals twice the incoming one in the usual
backscattering configuration, to an excellent approximation. In the presence
of polaritons, two incoming and two outgoing branches have to be considered
and acoustic phonons with up to four different wavevectors can be
simultaneously involved in the scattering process \cite{weisbuch1982}.
Moreover, the wavevectors and thus the Brillouin shifts directly reflect the
polariton dispersion. Experimental observations of such features are
considered as one of the most direct evidence of the existence of excitonic
polaritons in bulk semiconductors.

It has been demonstrated \cite{andreani1994}\ that the previous analysis can
be extended, using an effective $\beta$, to superlattices in the long
wavelength limit, which applies to the optical properties studied here.
Moreover a more realistic expression for the polariton is obtained when the
homogeneous broadening of the exciton due to its finite lifetime is included:%

\begin{equation}
\frac{c^{2}k^{2}}{\omega^{2}}=\varepsilon_{0}+\frac{4\pi\beta\omega_{X}^{2}%
}{\omega_{X}^{2}-\omega^{2}-i\gamma\omega} \label{eq}%
\end{equation}
The combination of damping and dispersion in the exciton part of the polariton
leads to a complex situation which has not been described consistently as
regards its consequences on Brillouin scattering. However the case of a
discrete non dispersive exciton appears to be a simple model system whose
experimental realization has never been considered up to now. In this case, a
single polariton branch exists at any energy and its dispersion and damping
can be described with a dispersive complex dielectric constant $\varepsilon
(\omega)$ equal to the right hand side of \ Eq.(\ref{eq}). We illustrate in
Fig. 1 the Brillouin scattering mechanism in this situation for two cases: in
the left panel, the low energy Brillouin line LA with dispersion $\omega=vk$
($v$ is the sound velocity) is involved and the incoming and outgoing
polaritons take place at very close points in the dispersion. When the energy
is in close resonance with the exciton, both the incoming and the outgoing
channels contribute simultaneously to the resonance. In the right panel, the
lowest folded acoustic line FLA-1 with dispersion $\omega=v(\frac{2\pi}{d}-k)$
($d$ is the period of the superlattice) is involved and the constant energy
offset due to folding gives a unique possibility to separate the incoming and
outgoing photon contributions. In this figure we used a damping of 0.3 meV, a
value corresponding to the experiments discussed below. This damping is
significantly larger than the longitudinal transverse splitting, thus letting
the LT splitting irrelevant, but much smaller than the Rabi gap, leading to
significant modification of the dielectric constant around the exciton energy
due to the polariton coupling, which is the topics of the experimental study
presented in this letter.

We have performed resonant Brillouin scattering on a multi-quantum well with
40 periods with 17.1 nm of GaAs and 7.5 nm of AlAs in each period. The
Brillouin experiments have been performed with a Coherent MBR single-mode
Ti-Sa tunable ring laser with wavelength stability better than 0.01 cm$^{-1}$.
Resonance profiles have been measured close to different confined exciton
levels and we focus here on the E1-HH1 transition between the lowest
conduction electron state and the one of the heavy hole. The resonance curve
has been followed using laser energy steps close to 25
%TCIMACRO{\U{b5}}%
%BeginExpansion
$\mu$%
%EndExpansion
eV (0.2 cm$^{-1}$). The light emitted by the sample, including excitonic
luminescence, Brillouin scattering and the peak at the laser energy, has been
dispersed by a Dilor XY triple Raman spectrometer in additive mode and
recorded with a nitrogen cooled CCD multichannel detector. Experimental
resolution is close to 0.2 cm$^{-1}$. We discuss in this letter experimental
results obtained at 80K. At higher temperatures, all intensities decrease,
thus reducing the accuracy of the measurements. At lower temperature, the
excitonic luminescence increases strongly and dominates over the Brillouin
spectra below 20K. Thanks to the large intensity of the Brillouin lines in the
intermediate temperature range (50-80K), we have been able to record in the
same CCD window the different scattered lines without having to filter out the
laser line or the exciton luminescence. This allows a very accurate
determination of the energy shifts as a function of the incident laser
energy.
%TCIMACRO{\FRAME{ftbpFU}{3.5812in}{2.7536in}{0pt}{\Qcb{Left panel: resonant
%Brillouin spectra at four different incident laser energies across the E1HH1
%excitonic resonance. In the right panel, the two curves with the smallest and
%the largest incident energies in the left panel are shown with an energy scale
%relative to the laser position (Raman shift). The vertical lines point out the
%change in the Raman shifts across the resonance.}}{}{Figure}%
%maintain-aspect-ratio TRUE;  display "USEDEF";  valid_file "T";
%width 3.5812in;  height 2.7536in;  depth 0pt;  original-width 4.2393in;
%original-height 3.2517in;  cropleft "0";  croptop "1";  cropright "1";
%cropbottom "0";  tempfilename 'LKF2ZC05.wmf';tempfile-properties "XPR";}}}%
%BeginExpansion
\begin{figure}
[ptb]
\begin{center}

\includegraphics[
natheight=2.753600in,
natwidth=3.581200in,
height=2.7536in,
width=3.5812in
]%
{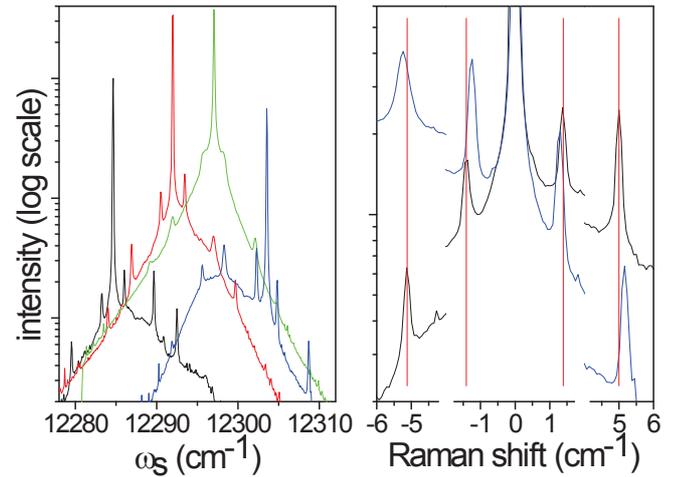}%
\caption{Left panel: resonant Brillouin spectra at four different incident
laser energies across the E1HH1 excitonic resonance. In the right panel, the
two curves with the smallest and the largest incident energies in the left
panel are shown with an energy scale relative to the laser position (Raman
shift). The vertical lines point out the change in the Raman shifts across the
resonance.}%
\end{center}
\end{figure}
%EndExpansion

We show in Figure 2 Brillouin spectra measured at a few different incident
energies across the excitonic transition. The Brillouin line, with a shift
close to 1.3 cm$^{-1}$, and the lowest folded lines, with shifts around 5 and
8 cm$^{-1}$, are recorded on each side of the laser line, respectively
corresponding to Stokes and anti-Stokes processes, and on the top of the
excitonic luminescence. In the left panel, one clearly sees the large
variations of the intensity, plotted in logarithmic scale, and of the
linewidth of the different peaks. In the spectrum in green, with laser energy
close to the exciton transition energy, all peaks are significantly broadened.
In the spectra in blue and red, the folded lines with absolute energies close
to the exciton transition are significantly broadened as well. When plotted
against the laser energy (right panel), a variation of the line shifts becomes
apparent, going towards opposite directions for the Brillouin line and the
lowest folded line respectively. We have performed a detailed line shape
analysis in order to extract in a systematic way the positions and the widths
of the different lines in the spectra, after having removed the background due
to the excitonic luminescence. Thanks to the coexistence of the laser line and
the Brillouin lines in the same spectra, a very high accuracy has been
obtained for the Brillouin shift, the distance between the inelastic lines and
the elastic one.%

%TCIMACRO{\FRAME{ftbpFU}{3.1886in}{4.2099in}{0pt}{\Qcb{Left panels: Variation
%with the incident energy of the Brillouin shift (full squares) and the
%Brillouin line width (open circles) deduced from the experimental spectra: LA
%phonon (lower panel), folded phonons FLA-1 (middle panel) and FLA+1 (upper
%panel). In each panel, the data are represented for the Stokes (red labels)
%and the Anti-Stokes (black labels) lines. Right panels: calculated values of
%the same quantities as in the left panel according to the model described in
%the text. The vertical line in all panels indicates the fitted exciton energy
%(incoming resonance).}}{}{Figure}{\special{ language "Scientific Word";
%type "GRAPHIC";  maintain-aspect-ratio TRUE;  display "USEDEF";
%valid_file "T";  width 3.1886in;  height 4.2099in;  depth 0pt;
%original-width 3.1445in;  original-height 4.5602in;  cropleft "0";
%croptop "0.9562";  cropright "1";  cropbottom "0.0437";
%tempfilename 'LKEK6R02.wmf';tempfile-properties "XPR";}}}%
%BeginExpansion
\begin{figure}
[ptb]
\begin{center}

\includegraphics[
natheight=4.209900in,
natwidth=3.188600in,
height=4.2099in,
width=3.1886in
]%
{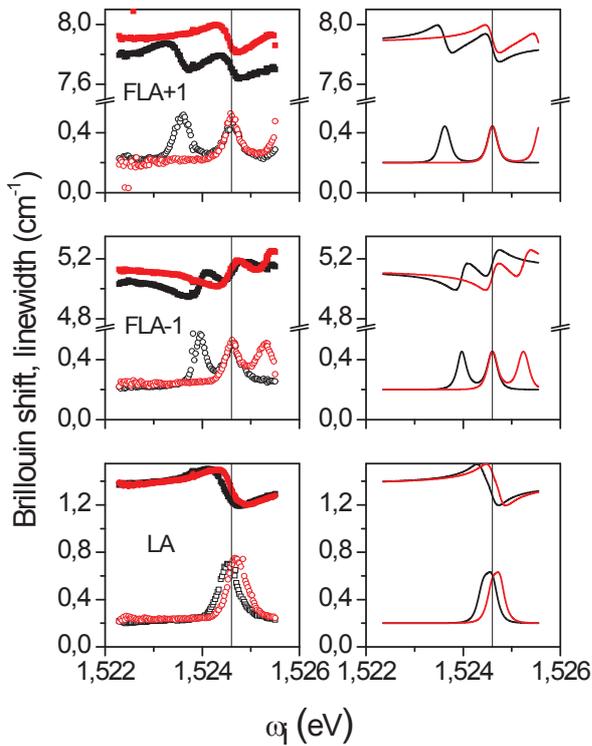}%
\caption{Left panels: Variation with the incident energy of the Brillouin
shift (full squares) and the Brillouin line width (open circles) deduced from
the experimental spectra: LA phonon (lower panel), folded phonons FLA-1
(middle panel) and FLA+1 (upper panel). In each panel, the data are
represented for the Stokes (red labels) and the Anti-Stokes (black labels)
lines. Right panels: calculated values of the same quantities as in the left
panel according to the model described in the text. The vertical line in all
panels indicates the fitted exciton energy (incoming resonance).}%
\end{center}
\end{figure}
%EndExpansion

We have deduced very systematic variations as a function of the incident
energy, with a very high signal to noise ratio. As shown in Fig. 3, these
variations exhibit systematic trends when the different lines and the Stokes
and Anti-Stokes components for each one are compared. The main feature is an
oscillation of the shift, with typical amplitude of 0.2 cm$^{-1}$, and a peak
in the width close to the exciton resonance. Outside the resonance the
linewidth is dominated by the resolution of the spectrometer amounting to 0.2
cm$^{-1}$ while its peak value is of the order of 0.5 cm$^{-1}$. One should
note that the intrinsic acoustic phonon lifetime is extremely long and has
negligible influence on the linewidth. In the case of the folded modes, two
lineshift oscillations are observed both in the Stokes and the Anti-Stokes
components. One of these oscillations is common to both components while the
second one appears at higher (resp. lower) incident energy for the Stokes
(resp. Anti-Stokes) component. We attribute these two structures to the
incoming and outgoing resonances. At the incoming resonance, when the laser
energy coincides with the exciton, one expects a common feature in all the
experimental traces, which is the case in our data for 1.5246 eV. The outgoing
resonance is shifted from the incoming one by the Brillouin shift, positive or
negative whenever the Stokes or the Anti-Stokes component is considered. This
is verified in the experimental data. In the case of the LA line, the
Brillouin shift is too small to allow the separation of the two resonance
channels and a single resonance is observed at a slightly different position
for Stokes and the Anti-Stokes components. The amplitude of the oscillation
and the linewidth peak value are larger than for the folded lines because they
reflect the combination of two resonances. To summarize, the excitonic picture
as schematically represented in Fig.1 gives an excellent qualitative
description of the experimental features. In particular, it reproduces the
observation of three identical oscillations in the energy shift and three
identical peaks in the line broadening, one being common to the Stokes and the
Anti-Stokes components and the other ones symmetrically shifted towards higher
or lower energy in ST and AS respectively. We will now show that a theoretical
model based on the discrete damped polariton dispersion allows to
quantitatively well fit the main experimental features based on the expected
background dielectric constant and exciton energy and provides a unique access
to the exciton homogeneous broadening.

Based on the mechanisms presented in Fig. 1, it is straightforward to
calculate the Brillouin shifts which fulfill both the conservation of the
energy and the wavevector in a scattering event with a LA or a FLA line, both
in the Stokes and Anti-Stokes channels. The result of this calculation is
shown in the right panels of Fig.3, with the same colors as used in the left
panel for experimental results. The full width at half maximum is obtained
from the expression $2v(Im(k_{i})+Im(k_{s}))$ in which $k_{i}$ and $k_{s}$ are
the wavevectors of the incident and scattered polaritons. The experimental
resolution (0.2 cm$^{-1}$) is taken account by a convolution formula strictly
valid for Gaussian profiles only. We have taken the parameters given for bulk
GaAs in Ref.\cite{weisbuch1982} : a background dielectric constant equal to
12.55 and an exciton cross section corresponding to longitudinal transverse
splitting of 0.086 meV. The exciton energy has been fitted to 1.5246 eV and
the lifetime parameter $\gamma$ to $0.30\pm0.03$ meV. With these parameters,
the width and the amplitude of the structures in the energy and linewidth
resonance profiles are very well reproduced. The peak linewidth is slightly
underestimated by an amount within the uncertainty in the lineshape fitting
and the simplified treatment of the instrumental broadening. The bulk $\beta$
value allows a good description of the experimental results as the varaition
due to confinement is expected to be small for such a wide quantum well
\cite{andreani90} Some discrepancy remains in the background variations of the
line energies, negligible for the Brillouin line and of increasing magnitude
with the energy shift of the line. This is likely to be due to additional
contributions to the background dielectric constant, f. i. due to higher
energy band to band transitions, not included in the model.

We would like to point out finally that the question of light scattering
intensity resonances in the polariton frame has been much debated
\cite{bendow1978} and the conclusion remained unclear whether including the
polariton picture significantly changes the resonance curves in bulk material,
as compared to the exciton one. The analysis of our intensity measurements is
a topics in itself and we leave it for a forthcoming publication together with
additional results on resonances close to the E2HH2 and E3HH3 excitons.\ 

In conclusion, we have observed very clear evidences of the polaritonic
coupling in light scattering by acoustic and folded acoustic phonons in a
multi-quantum-well when the polariton damping is larger than the longitudinal
transverse splitting but significantly smaller that the Rabi
splitting.\ Thanks to the combination of a specifically designed high
resolution set up, a multi-quantum well with much thicker layers than in
previous studies and an optimization of the experimental temperature,
exceptionnaly clear variations of both position and width of Brillouin lines
have been demonstrated \ for the first time when the photon energies satisfies
either the incoming or the outgoing resonance. Additional novel results are
the observation of similar consistent effects on acoustic folded lines. A
quantitative description is obtained in a damped polariton picture provided a
broadening of 0.3 meV is included,\ a value significantly smaller than the
exciton linewidth measured in photoluminescence excitation. This suggests that
light scattering gives access to new information on the damping mechanisms in
multi-quantum wells, possibly excluding the inhomogeneous broadening from well
to well owing to the delocalized nature of the probing acoustic phonons.

This work is a part of the Internatioanl French Argentinian Nanoscience
Laboratory, LIFAN.\ We would like to thank very much J.\ Bloch for several
discussions on the polariton properties.


\begin{thebibliography}{9}                                                                                                %


\bibitem {weisbuch1982}C. Weisbuch and R.G. Ulbrich, in Light Scattering in
Solids III, edited by M. Cardona and G. G\"{u}ntherodt (Springer, Berlin
Heidelberg New York, 1982) p. 207

\bibitem {bendow1978}B.\ Bendow, in Electronic Structure of Noble Metals and
Polariton-Mediated Light Scattering, (Springer, Berlin \ Heidelberg, New York,
1978) p.69

\bibitem {weisbuch1992}C. Weisbuch, M. Nishioka, A. Ishikawa, and Y. Arakawa,
Phys. Rev. Lett. \textbf{69}, 3314 (1992)

\bibitem {fainstein1997}A. Fainstein, B. Jusserand, and V. Thierry-Mieg, Phys.
Rev. Lett. \textbf{78}, 1576 (1997)

\bibitem {jusserand1989}B. Jusserand and M. Cardona, in Light Scattering in
Solids V, edited by M. Cardona and G. G\"{u}ntherodt (Springer, Berlin
Heidelberg New York, 1989) p. 49

\bibitem {ruf1993}T. Ruf, V. I. Belitsky, J. Spitzer, V. F. Sapega, M.
Cardona, and K. Ploog, Phys. Rev. Lett. \textbf{71}, 3035 (1993)

\bibitem {andreani1994}E.L.\ Ivchenko, Sov.\ Phys.\ Solid State 33, 1344
(1991); L.\ C.\ Andreani, Physics Letters A \textbf{192}, 99 (1994)

\bibitem {hopfield1958}J. J. Hopfield, Phys. Rev. \textbf{112}, 1555 (1958)

\bibitem {andreani90}L.C.\ Andreani, and A.\ Pasquarello, Phys.\ Rev.\ B 42
8928 (1990)
\end{thebibliography}
\end{document}